\documentclass[nofootinbib,twocolumn]{revtex4-2}
\usepackage[T1]{fontenc}
\usepackage{graphicx,subcaption}
\usepackage{rotating}
\usepackage{bm}        
\usepackage{amssymb}   
\usepackage{float}
\usepackage{tikz}
\usepackage{diagbox}
\usepackage{braket}
\usepackage{color} 
\usepackage{colortbl}
\usepackage{amsmath}
\usepackage{xcolor}
\usepackage{soul}
\usepackage{float,subcaption}
\usepackage[rightcaption]{sidecap}
\DeclareCaptionJustification{justified}{\justifying}
\usepackage{booktabs}
\usepackage{siunitx}
\usepackage{caption}
\usepackage{multirow}

\makeatletter

    \def\CT@@do@color{%
      \global\let\CT@do@color\relax
            \@tempdima\wd\z@
            \advance\@tempdima\@tempdimb
            \advance\@tempdima\@tempdimc
    \advance\@tempdimb\tabcolsep
    \advance\@tempdimc\tabcolsep
    \advance\@tempdima2\tabcolsep
            \kern-\@tempdimb
            \leaders\vrule
                    \hskip\@tempdima\@plus  1fill
            \kern-\@tempdimc
            \hskip-\wd\z@ \@plus -1fill }
    \makeatother

\def\k1{k_1}
\def\k2{k_2}
\def\q1{q_1}
\def\q2{q_2}

\def\({\left (}
\def\){\right )}
\def\[{\left [}
\def\]{\right ]}

\newcommand{\beq}{\begin{equation}}
\newcommand{\eeq}{\end{equation}}

\DeclareMathAlphabet\mathbfcal{OMS}{cmsy}{b}{n}

\begin{document}

\title{Lattice stitching by eigenvector continuation for Holstein polaron
}
\author{Elham Torabian and Roman V. Krems}
 \affiliation{
Department of Chemistry, University of British Columbia, Vancouver, B.C. V6T 1Z1, Canada \\
Stewart Blusson Quantum Matter Institute, Vancouver, B.C. V6T 1Z4, Canada }

\date{\today}

\begin{abstract}
Simulations of lattice particle - phonon systems are fundamentally restricted by the exponential growth of the number of quantum states with the lattice size. 
Here, we demonstrate an algorithm that constructs the lowest eigenvalue and eigenvector for the Holstein model in extended lattices from eigenvalue problems for small, independent lattice segments. This leads to exponential reduction of the computational Hilbert space and allows applications of variational quantum algorithms to particle - phonon interactions in large lattices. We illustrate that the ground state of the 
Holstein polaron in the entire range of electron - phonon coupling, from weak to strong, and the lowest phonon frequency ($\omega/t = 0.1$) considered by numerical calculations to date can be obtained from a sequence of up to four-site problems. When combined with quantum algorithms, the present approach leads to a dramatic reduction of required quantum resources. We show that the ground state of the Holstein polaron in a lattice with 100 sites and 32 site phonons
can be computed by a variational quantum eigensolver with 11 qubits. 
\end{abstract}

\maketitle

Particle-phonon interactions are fundamental to thermal, mechanical, conducting and optical properties of materials, leading to polarons \cite{simon2013oxford, holstein1959studies} and bipolarons \cite{alexandrov1994bipolarons} and determining dynamics of excitons or excitations in photoexcited molecular crystals \cite{scholes2006excitons,park2009bulk,spano2010spectral,miyata2017large}, light-harvesting complexes \cite{scholes2011lessons,damjanovic2002excitons,ye2012excitonic,ferretti2016dark},  polaritonic chemistry \cite{feist2018polaritonic,xiang2020intermolecular}, impurities in ultracold gases \cite{schirotzek2009observation,koschorreck2012attractive,jorgensen2016observation,hu2016bose}, and in some more exotic physical settings \cite{baggioli2015electron,sous2020fractons1,sous2020fractons}. Many numerical approaches for accurate quantum calculations for models with phonon couplings have been developed, including variational exact diagonalization \cite{bonvca1999holstein,bonvca2007numerical,bonvca2021dynamic}, matrix product-state techniques \cite{jeckelmann1998density,zhang1999dynamical,dorfner2015real,kloss2019multiset}, variational Green's function calculations \cite{carbone2021numerically,kairon2023extrapolation,sous2017bipolarons,berciu2006green,berciu2007systematic,ebrahimnejad2014dynamics,lau2007single,ebrahimnejad2016differences,bieniasz2016green,goodvin2006green,bieniasz2016green,berciu2011few,bieniasz2017orbiton,bieniasz2017orbiton}, Monte Carlo (MC) methods \cite{prokof1998polaron,prokof2008fermi,mishchenko2000diagrammatic,titantah2001free,kornilovitch1998continuous}, and other variational methods \cite{la1996variational, brown1997variational, romero1999self, wang2020zero, yang2024benchmarking}. However, applications of these calculations to extended lattice systems and the regime of low phonon frequencies ($\omega \rightarrow 0$, adiabatic regime) remain challenging because of the exponential growth of the number of phonon states with the number of lattice sites and with $1/\omega$. This has also restricted applications of emerging quantum computing platforms for simulations of models with particle - phonon interactions. Here, we demonstrate a variational algorithm that constructs the lowest eigenvalue and eigenvector for particle - phonon systems in extended lattices from eigenvalue problems for small, independent lattice segments. 
We show that this approach can be combined with a variational quantum eigensolver (VQE) to compute the energy of the Holstein polarons \cite{holstein1959studies} with a small number of qubits. 

We consider a particle in a lattice with $N_s$ sites coupled locally to $N_p$ phonons per site. This leads to $d = N_s \times N_{p}^{N_s}$ coupled quantum states. For an ordered, one-dimensional lattice with $N_s = \infty$ and nearest-neighbour hopping ($t$), the problem is the Holstein model \cite{holstein1959studies}, one of the benchmark polaron models. The lowest phonon frequency previously considered in numerical calculations of the Holstein polaron energy was $\omega/t = 0.1$ \cite{kairon2023extrapolation,kuchinskii2024generalized,yam2020peierls,jeckelmann1998density, yang2024benchmarking}, demonstrating the challenges of the numerical analysis of particle - phonon systems in the adiabatic regime. We show that the same calculations can be performed with a series of diagonalizations of two-site and four-site Hamiltonians, followed by continuation of the resulting eigenvectors into the full Hilbert space, 
yielding accurate polaron energy at  $\omega/t = 0.1$ and for the entire range of particle-phonon coupling strength, from weak to strong coupling. 
Finally, we combine the present approach with the VQE algorithm \cite{VQE} and demonstrate that the lowest energy of the Holstein polaron model for a lattice with 100 sites and 32 phonons per site can be computed using 11 qubits, a reduction from 507 qubits required for VQE with the full Hamiltonian matrix. Previous applications of VQE to the Holstein and Hubbard-Holstein models were restricted to six lattice sites \cite{denner2023hybrid, li2023efficient,macridin2018digital}. 

The present work is based on eigenvector continuation (EC) \cite{ECbase} introduced for identifying the extreme eigenvalues of parametrized Hamiltonians by projections onto subspaces of eigenvectors of a set of fixed Hamiltonians. Initially developed to extrapolate quantum many-body wave functions in large vector spaces, EC has found applications in quantum MC methods to address the sign problem \cite{sarkar2021convergence, ECMC}, as a fast emulator for quantum many-body systems \cite{duguet2024colloquium, ECManyBody1,ekstrom2019global}, and as a resummation technique for perturbation theory \cite{demol2020improved}. EC has proven effective in constructing efficient emulators for shell-model calculations for valence spaces, parameter optimization, and uncertainty quantification \cite{yoshida2022constructing}, as well as for applications in quantum scattering \cite{zhang2022fast,drischler2021toward,melendez2021fast}. It has been applied to calculate three-nucleon forces \cite{wesolowski2021rigorous}, analyze interacting cold atom systems \cite{eklind2021eigenvector}, and compute excited states of anharmonic oscillators \cite{franzke2022excited}. The original formulation of EC was previously combined with VQE for quantum chemistry calculations as well as for spin models to simplify the requirements for quantum hardware \cite{schrader2022eigenvector,francis2022subspace,mejuto2023quantum}.
Here, we illustrate that EC can efficiently and accurately extrapolate from tensor products of Hilbert spaces into a full Hilbert space of a particle - phonon system, which allows for an exponential reduction of the complexity of variational quantum algorithms for polaron problems.

{\it Method.} The Hamiltonian considered in this work can most generally be written as (with $\hbar = 1$):
\begin{equation}
    \begin{split}
        H=& \; \epsilon \displaystyle\sum_{i}^{N_s} a_i^{\dagger} a_i +\displaystyle\sum_{i,j>i}^{N_s} t_{ij} (a_i^{\dagger} a_j+a_j^{\dagger} a_i)\\+
        & \omega \displaystyle\sum_{i}^{N_s} b_{i}^{\dagger} b_{i}+ g \displaystyle\sum_{i=1}^{N_s} \; a_{i}^{\dagger} a_{i} (b_{i}^{\dagger} + b_{i}),
    \end{split}
    \label{eq1}
\end{equation}
where $a_i^{\dagger}$ creates a  bare particle and $b^\dagger_i$ creates a phonon on site $i$,  and $t_{ij}$ is the amplitude for the bare particle hopping between sites $i$ and $j$. 
We consider the Holstein polaron model with an ordered one-dimensional lattice, $\epsilon = 0$ and $t_{ij} = t = 1$ restricted to nearest neighbour hopping. We quantify the strength of the particle - phonon coupling by 
\begin{equation}
    \lambda = g^2/2t\omega.
    \label{eq2}
\end{equation}
We compare the results of our calculations with the analytical solution for the strong electron - phonon interaction regime \cite{macridin2003phonons}, as well as quantum MC calculations \cite{macridin2003phonons} and the results of variational Green's function calculations \cite{carbone2021numerically} at arbitrary values of $\lambda$.

The objective of EC is to estimate an extreme eigenvalue corresponding to eigenvector $| v({\bm c}_t) \rangle$ of a parametrized Hamiltonian $\hat H ({\bm c}_t)$, where ${\bm c}_t$ denotes collectively the target parameters. 
Assuming that $\hat H ({\bm c}_t)$ cannot be directly diagonalized, EC aims to determine $| v({\bm c}_t) \rangle$ as a linear combination of $k$ exact ground-state eigenvectors $\ket{v({\bm c}_1)}, \dots , \ket{v({\bm c}_k)}$ of  $\hat H ({\bm c}_i)$, where ${\bm c}_i$ is a fixed sets of Hamliltonian parameters and  
the different sets $i \in [1, k]$ are the training points for EC. The target Hamiltonian $\hat H({\bm c}_t)$ is projected onto the subspace spanned by the training eigenvectors $\ket{v(c_1)}, \dots , \ket{v(c_k)}$, to yield the effective Hamiltonian matrix ${\bf H}_{\rm eff}$ of size $k \times k$ with elements: 
\begin{equation}
    {\bf H}_{{\rm eff},ij} ({\bm c}_t) = \braket{v({\bm c}_i)|\hat H({\bm c}_t)|v({\bm c}_j)}
    \label{eq3}
\end{equation}
Solving the generalized eigenvalue problem
\begin{equation}
    {\bf H}_{\rm eff}({\bm c}_t) |{v}({\bm c}_t)\rangle = {E}  {\bf S} |{v}({\bm c}_t) \rangle
    \label{eq4}
\end{equation}
with the elements of matrix $\bf S$ given by $\braket{v({\bm c}_i)|v({\bm c}_j)}$, yields the approximation of the ground-state energy $E$ and the corresponding eigenvector of the target Hamiltonian.  

In the present work, we consider ${\bm c}_t$ to comprise all $t_{ij}$ in Eq. (\ref{eq1}). The training points are obtained by completely decoupling the lattice by setting a subset of particle hopping amplitudes to zero. 
To obtain an accurate representation of the ground state of the entire lattice with $N_s$ sites, we use the following algorithm. 
We first break the lattice into $N_s/N_k$ equal segments, where $N_k \ll N_s$. In the following calculations, $N_k = 2, 3$ or $4$.
The target lattice is thus divided into $k$ independent fragments each with $N_k$ sites. Given the translational symmetry, the entire set of the training vectors $|v ({\bm c}_i) \rangle~\forall~i$
can be constructed by a single exact diagonalization of the Hamiltonian matrix with the size $d_k \times d_k$, where  $d_k = N_k \times N_{p}^{N_k}$.  The polaron state is then obtained by 
diagonalization of $ {\bf H}_{\rm eff}$ of size $k \times k$.   We find that the accuracy of the calculations can be significantly improved by including additional segments (more training points) corresponding to all possible overlaps between the uncoupled segments, with each overlap comprising $N_k$ sites. This can also be viewed as representing the lattice of $N_s$ sites by all possible $N_k$-site fragments.

The current approach thus aims to represent the ground state of the full Hamiltonian by a superposition of vectors, each living in the tensor product of the Hilbert spaces of small lattice segments. This makes this approach particularly well-suited for reducing the complexity of variational quantum algorithms. In VQE, a quantum computer calculates the expectation value $\langle \hat H \rangle$ of a Hamiltonian $\hat H$ over a parameterized quantum state $|{\psi(\bm\theta)}\rangle$. The value $\langle \hat H \rangle$ is then minimized with respect to $\bm\theta$ to yield the lowest eigenvalue of $\hat H$. Following Refs. \cite{kandala2017hardware,asnaashari2023compact}, we use the ansatz 

\begin{equation}
    \ket{\psi(\bm\theta)} = \prod_{d=0}^{p-1}\left[\prod_{q=0}^{n-1} U^{q,d}(\theta^q_{d})\times U^d_{\rm ent}\right] \times \prod_{q=0}^{n-1} U^{q, p}(\varphi^q_{k})\ket{0}^{\otimes n},  
    \label{eq5}
\end{equation}
where $U^{q,d}(\theta)$ represent single-qubit rotations $R_Y = \exp(-i\theta\sigma_Y/2)$ with Pauli matrix $\sigma_Y$ acting on qubit $q$, $p$ is the number of repetitions of the rotation and entanglement blocks, and $U^d_{\rm ent} = \prod_{q=0}^{n-1}{\rm CNOT}(q, q+1)$ entangles $n$ qubits by CNOT gates. 

The Hamiltonian matrix $\bf H$ is expanded in Pauli strings as follows: 
\begin{equation}
    {\bf H} = \sum_{i = 0}^{4^n} A_i K^i_1\otimes K^i_2 \ldots \otimes K^i_n 
    \label{eq6}
\end{equation}

with $K^i_j \in \{\sigma_X, \sigma_Y, \sigma_Z, I\}$ acting on qubit $j$, and
\begin{equation}
    A_i = \frac{1}{2^n}{\rm Tr}[(K^i_1\otimes K^i_2 \ldots \otimes K^i_n)\cdot {\bf H}].
    \label{eq7}
\end{equation}
This expansion scales exponentially with the number of qubits $n$. However, the decoupling of the lattice Hamiltonian by EC as introduced above reduces the scaling to $4^{n_{\rm EC}}$, with $n_{\rm EC}$ given by 
\begin{equation}
    n_{\rm EC} = \left ( \log N_k + N_k \log N_p \right )/ \log 2.
    \label{eq8}
\end{equation}
Given that $N_k$ is a small fixed number (2, 3, and 4 in the present work), this provides a dramatic reduction of the VQE computation complexity. 
More specifically, the reduction of the number of required qubits is 
\begin{equation}
    \frac{n_{\rm EC}}{n} \approx \frac{N_k}{N_s} 
    \label{eq9}
\end{equation}
We refer to the algorithm based on a combination of EC and VQE as EC-VQE.

 \begin{figure}[hbt!]
    \centering
    \includegraphics[width=\linewidth]{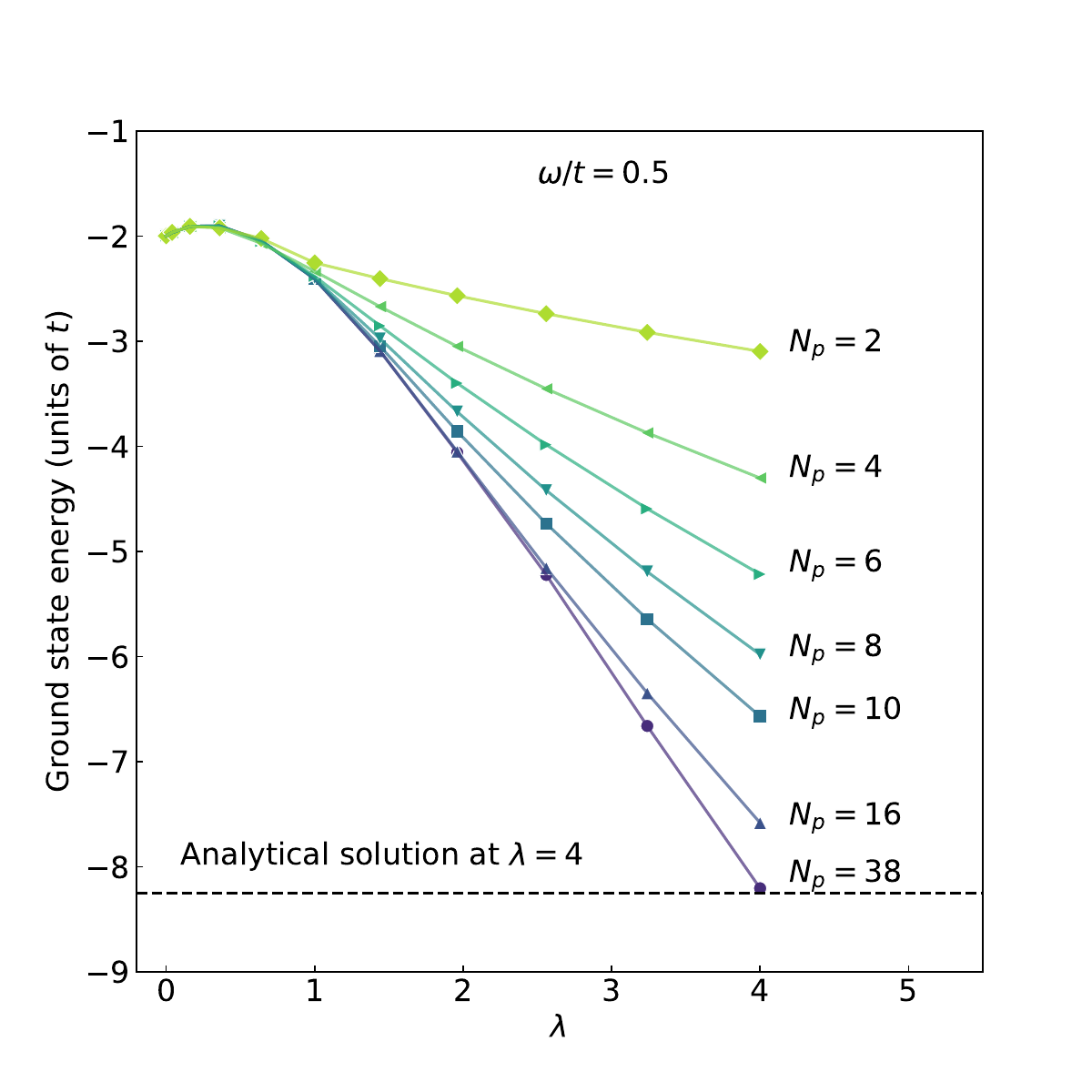}
    \caption{Ground state energy of the Holstein polaron for a lattice with  $N_s=100$ sites, $\omega/t = 0.5$ and $N_p$ phonons per site computed with the lattice stitching algorithm described in the text. The horizontal dashed line represents the third-order analytical approximation at $\lambda = 4$  \cite{marchand2011polaron}. The lowest energy result represented by circles is obtained by diagonalizing a matrix of size   $2888 \times 2888$ followed by the diagonalization of a matrix of size $50 \times 50$.
    }
    \label{Fig1}
\end{figure}

\begin{figure}[hbt!]
    \centering
    \includegraphics[width=\linewidth]{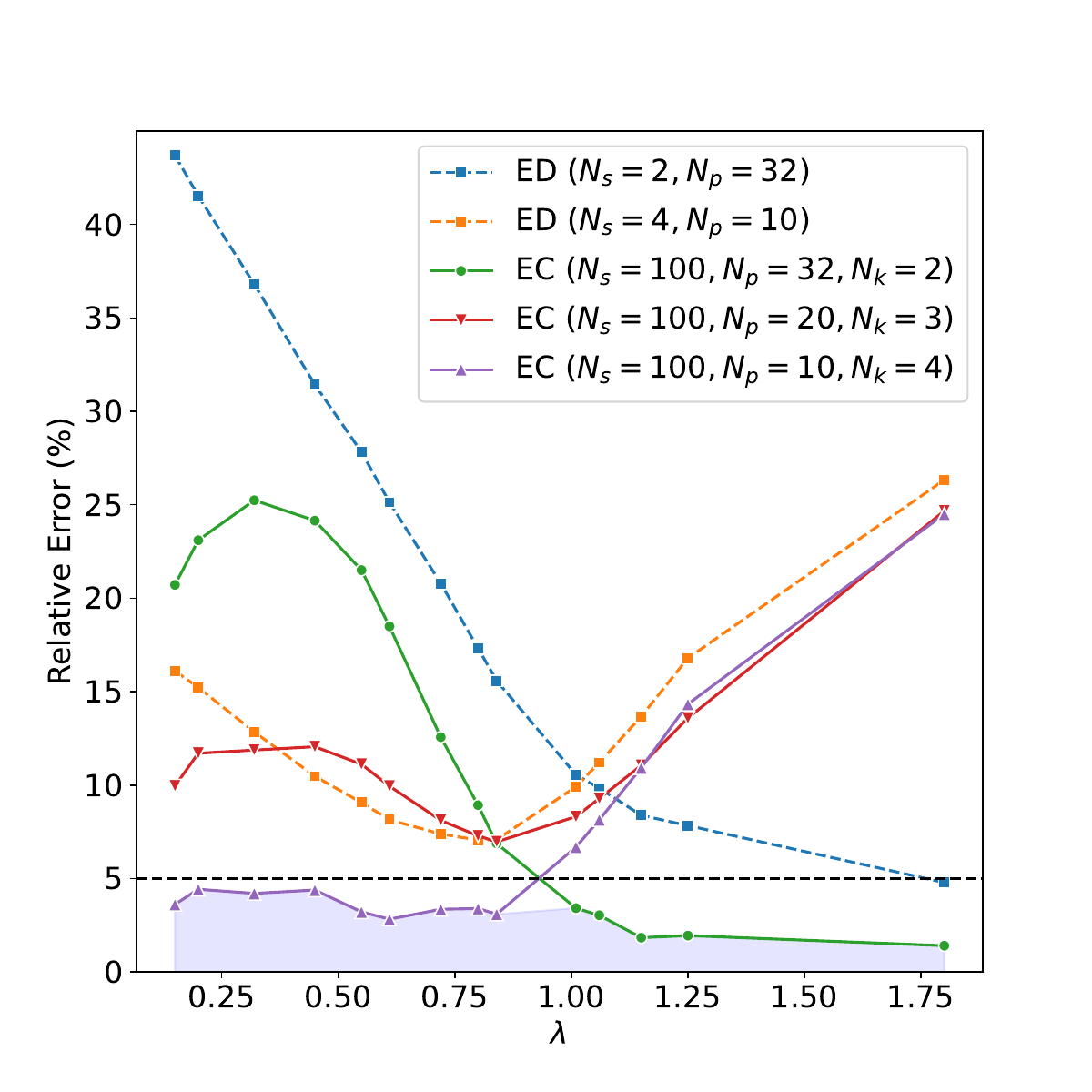}
    \caption{
    The relative error (represented by the shaded area) of the EC calculation of the Holstein polaron energy for a lattice with  $N_s=100$ sites and $\omega/t = 0.1$ as a function of $\lambda$. The dashed lines show the lowest eigenvalue for a single training point (isolated lattice segment) with two sites (blue squares) and four sites (orange squares). The solid lines are obtained using EC stitching of lattice segments with two sites (green circles), three sites (red downward triangles), and four sites (purple upward triangles). All errors are computed with respect to the polaron energy in Refs.  \cite{macridin2003phonons,carbone2021numerically}. }
    \label{Fig3}
\end{figure}

\begin{figure}[hbt!]
    \centering
    \includegraphics[width=\linewidth]{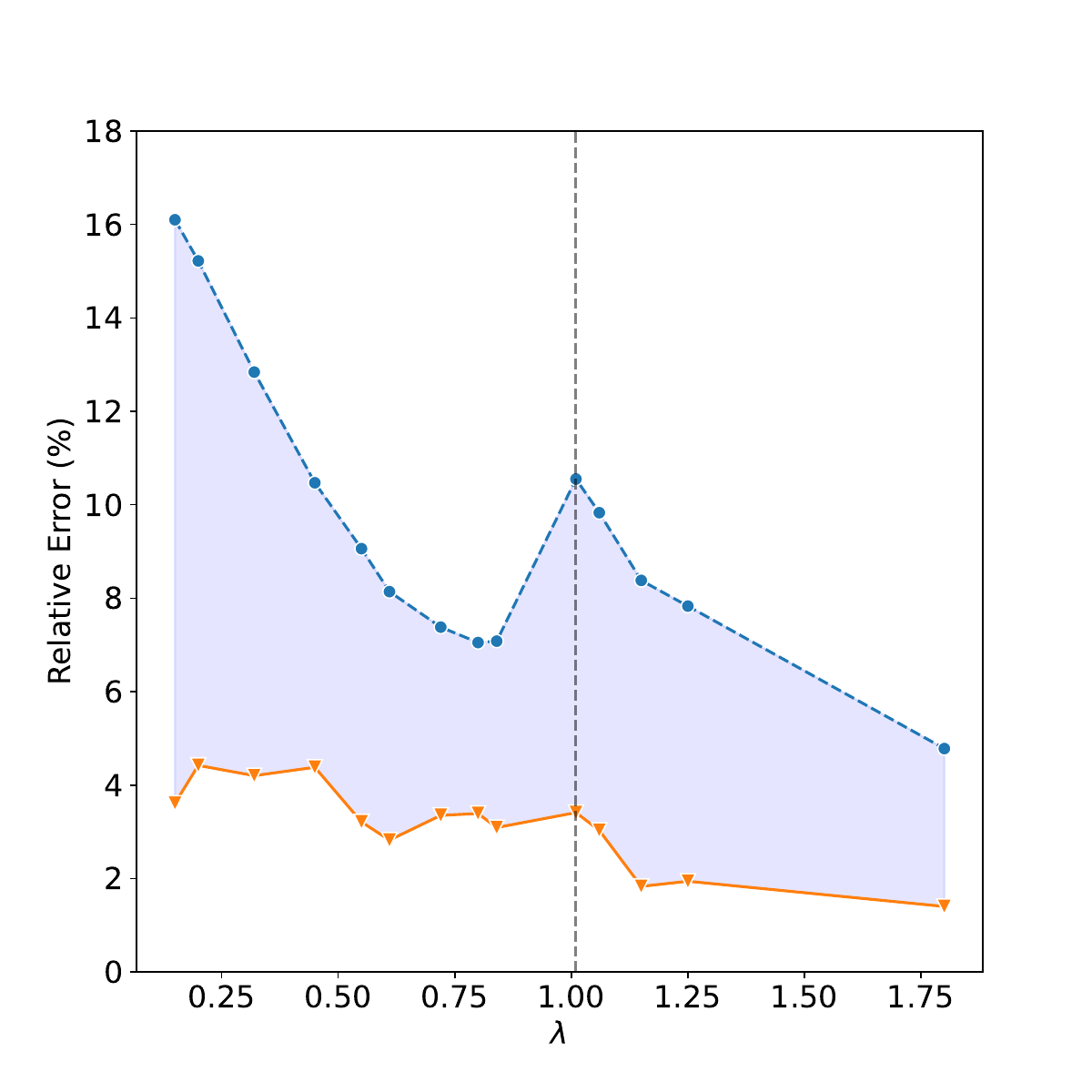}
    \caption{
    The improvement of the accuracy (represented by the shaded area) of the calculation of the Holstein polaron energy for a lattice with  $N_s=100$ and $\omega/t = 0.1$ 
    by lattice stitching using EC.  The blue circles show the results for the lowest eigenvalue for a single training point (isolated short lattice segment). The orange triangles are obtained using EC stitching of two-site fragments (at $\lambda \geq 1$) and four-site fragments (at $\lambda < 1$), augmented by overlap segments. All errors are computed with respect to the polaron energy in Refs.  \cite{macridin2003phonons,carbone2021numerically}.}
    \label{Fig4}
\end{figure}

\begin{figure}[hbt!]
    \centering
    \includegraphics[width=\linewidth]{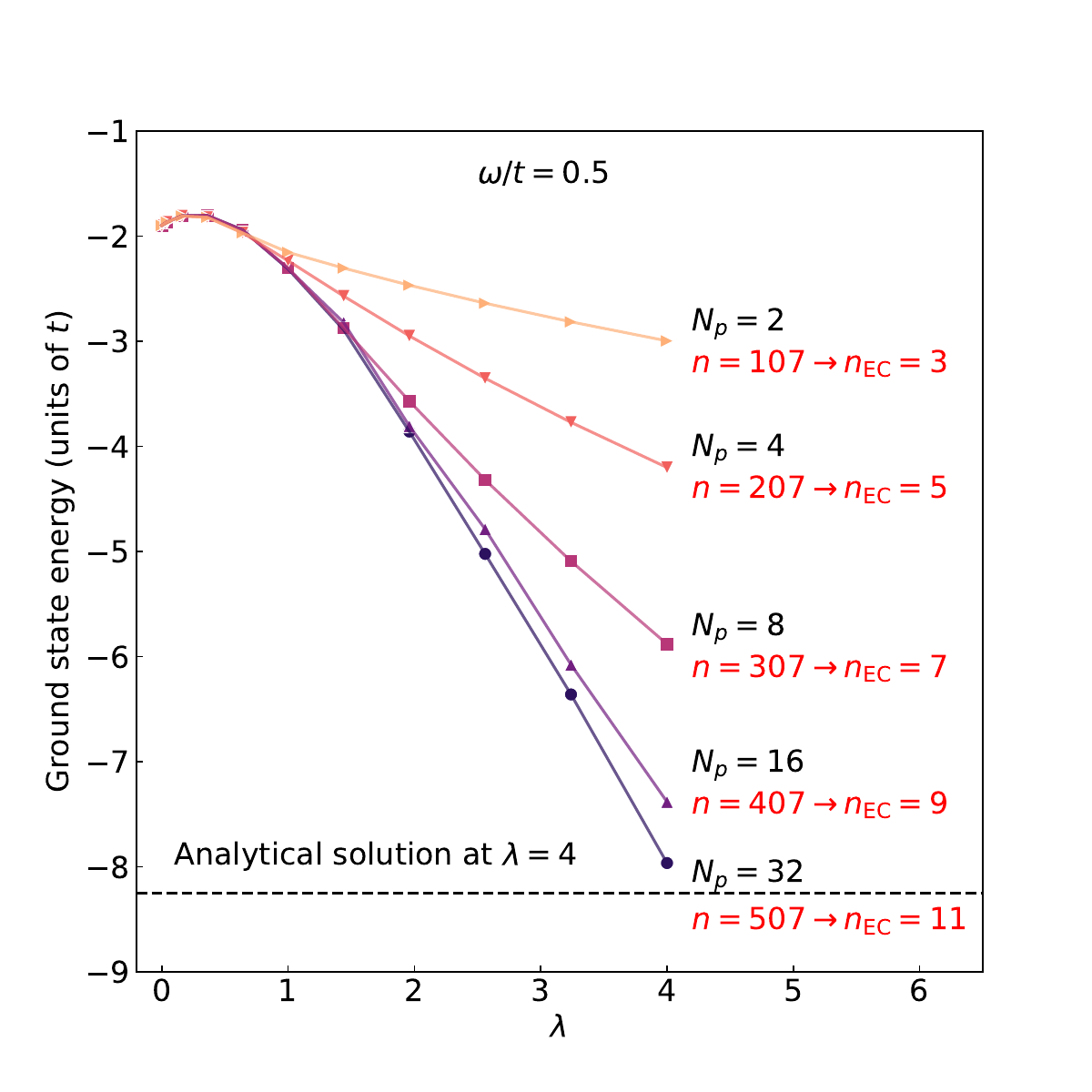}
    \caption{Ground state energy of the Holstein polaron model for a lattice with  $N_s=100$ sites, $\omega/t = 0.5$ and $N_p$ phonons per site computed by VQE combined with the lattice stitching algorithm described in text. The horizontal dashed line represents the third-order analytical approximation of ground state energy of Holstein polaron for $\lambda = 4$  \cite{marchand2011polaron}. The $n \rightarrow n_{\rm EC}$ notation indicates the reduction in the required number of qubits from VQE to EC-VQE. }
    \label{Fig6}
\end{figure}

\begin{figure}[hbt!]
    \centering
    \includegraphics[width=\linewidth]{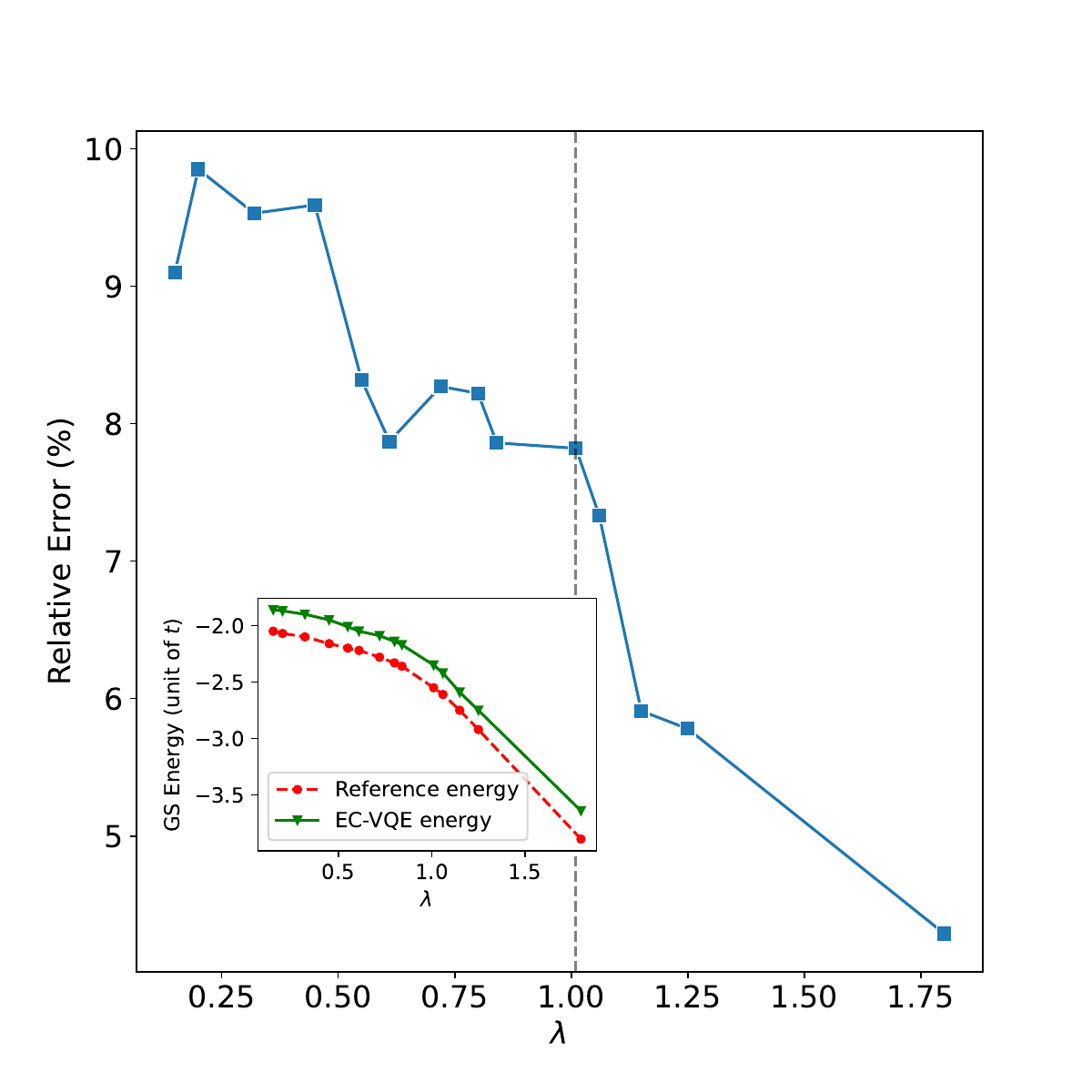}
    \caption{
    The relative error of the EC-VQE calculations of the Holstein polaron energy for a lattice with  $N_s=100$ sites, $\omega/t = 0.1$ and up to $N_p = 32$ phonons per site as a function of $\lambda$. These results are obtained using EC stitching of lattice segments with two sites (at $\lambda \geq 1$), and four sites (at $\lambda < 1$). All errors are computed with respect to the polaron energy in Refs.  \cite{macridin2003phonons,carbone2021numerically}.
    The inset shows the results of the EC-VQE calculations (green triangles) compared with the polaron energies (red circles) from Refs. \cite{carbone2021numerically,macridin2003phonons}.}
    \label{Fig7}
\end{figure}

{\it Results.} The ground-state energy of the Holstein polaron at strong coupling ($\lambda \gg 1$) is given by \cite{marchand2011polaron}
\begin{eqnarray}
E = \epsilon-2\lambda t-\frac{t}{\lambda}-2t \exp{(-2\lambda t/\omega)}
\end{eqnarray}
Figure \ref{Fig1} examines the convergence of the present EC approach with the number of phonons to this energy at $\lambda = 4$. The calculations are for the lattice with $N_s = 100$ sites and $\omega/t = 0.5$. 
In order to obtain the training points for these calculations, the lattice is split into two-site segments ($N_k = 2$). The lattice thus split is appended by segment overlaps, i.e. lattice sites 1--4 are represented by the following segments: (1,2), (3,4) and (2,3).
The results in Figure \ref{Fig1} converge to the Holstein polaron energy at $\lambda = 4$ with $N_p = 38$ phonons per site. The full Hamiltonian dimension for such a system (100 sites, 38 phonons per site) is $\approx 10^{160}$. Our results shown by circles involve the diagonalization of a matrix of size $2888 \times 2888$ followed by the diagonalization of a matrix of size $50 \times 50$. The relative error of our converged calculation at $\lambda = 4$ is $2\%$.

We now examine EC across the Holstein model parameters space. We calculate the Holstein polaron energy for $\omega/t = 0.1$ at different $\lambda$ values, from $\lambda=0.15$ to $\lambda = 1.8$. Figure \ref{Fig3} shows the relative error of the present approach for the lattice with $N_s = 100$ sites. This error represents the deviation of the polaron energy computed with the current approach from the results of the diagrammatic MC calculations \cite{macridin2003phonons} and generalized Green's function cluster expansions (GGCE) calculations   \cite{carbone2021numerically}.
The results illustrate that the ground-state Holstein polaron energy at $\omega/t = 0.1$ can be obtained within 5 \% of the reference energy for any $\lambda$.
Figure \ref{Fig4} quantifies the improvement obtained by applying EC to the results of corresponding single segment diagonalization. It should be mentioned that at $ \lambda \geq 1 $, it is sufficient to train EC with two-site fragments, as follows from Fig. \ref{Fig3}.

The results of Figs. \ref{Fig6} and \ref{Fig7} demonstrate how the EC approach introduced here reduces the complexity of quantum algorithms for computations of polaron properties. Figure \ref{Fig6} presents the same results as in Fig. \ref{Fig1}, but with the lowest eigenvalue of the Hamiltonian determined by EC-VQE through optimization of the parameters in Eq. (\ref{eq5}). 
For each number of lattice phonons $N_p$ considered, Fig. \ref{Fig6} illustrates the reduction of the number of required qubits for VQE by the current EC approach. 
In particular, the fully converged calculation at $\lambda = 4$, which requires $N_p = 32$ phonons,  can be performed by EC-VQE with 11 qubits, whereas the direct VQE scheme requires 507 qubits. 
The relative error of the EC-VQE calculation for the lattice with $N_s=100$ sites and $N_p = 32$ phonons per site is $5\%$. We use the ansatz (\ref{eq5}) with $p =1$, which requires optimization of $11$ parameters, yielding the optimal angles of the $R_Y$ gates.

Figure \ref{Fig7} shows the relative errors of EC-VQE, illustrating the deviation of the EC-VQE results from the reference polaron energies computed in Refs. \cite{carbone2021numerically, macridin2003phonons}. The results show that EC-VQE computations produce the polaron energy within 10 \% at all values of $\lambda \in [0.1, 2]$, covering the regimes of weak and strong particle - phonon coupling.
The inset of Fig. \ref{Fig7} shows the polaron energies calculated for the lattice with $N_s=100$ sites and up to $N_p = 18$ phonons per site. We use the ansatz (\ref{eq5}) with $p =2$, which requires optimization of up to $22$ angles of the $R_Y$ gates.

In summary, we have shown that accurate calculations of ground-state energy for lattice systems with Holstein particle - phonon couplings can be performed accurately and efficiently, using a sequence of eigenvalue problems for small lattice segments. 
The eigenvectors of these segments can be used to project the lowest-energy eigenstate of the full Hamiltonian onto a very restricted subspace without significant loss of accuracy for the ground-state energy. We illustrate this by a very efficient computation of the energy of the Holstein polaron deeply in the adiabatic regime, which yields an excellent agreement with literature results. The proposed algorithm is particularly well suited for reducing the complexity of variational quantum algorithms, which we demonstrate by combining the present EC approach with VQE for particle - phonon systems with 100 lattice sites and up to 32 phonons per site.  The present approach reduced the number of required qubits for VQE from 507 to 11. 

The authors thank Prof. Timur V. Tscherbul for bringing EC to our attention and Prof. Mona Berciu for identifying an error in our earlier calculation. This work was supported by NSERC of Canada.

\bibliography{References.bib}

\end{document}